%% file: hortop4.tex
\documentclass[a4paper,12pt]{article}
\usepackage{amsmath}
\usepackage{amssymb}
\usepackage{amsthm}
\usepackage{latexsym}
\usepackage{mathrsfs}
\usepackage[dvips]{graphicx}
\usepackage[colorlinks=true, citecolor=blue]{hyperref}
\usepackage{pstricks}
\usepackage{color}
\usepackage{tikz}
\usetikzlibrary{decorations.pathreplacing}
\def\f(#1){{\mathop{f}^{(#1)}}}
\def\m(#1){{\mathop{m}^{(#1)}}}
\def\C(#1){{\mathop{C}^{(#1)}}}
\def\p(#1){{\mathop{p}^{(#1)}}}

\def\ben{\begin{equation}}
\def\een{\end{equation}}
\def\bena{\begin{eqnarray}}
\def\eena{\end{eqnarray}}

\def\C{{\cal C}}

\def\mr{{\mathbb R}}

\newcommand{\mn}{{\mathbb N}}
\newcommand{\mz}{{\mathbb Z}}

\newcommand{\R}{{\mathcal R}}

\newcommand{\e}{{\rm e}}

\theoremstyle{theorem}

\newtheorem{res}{Result}

\begin{document}

\title{Further restrictions on the topology of stationary black holes in five dimensions}

\author{
Stefan Hollands$^{1,2}$\thanks{\tt HollandsS@Cardiff.ac.uk}\:
Jan Holland$^{1}$\thanks{\tt HollandJW1@Cardiff.ac.uk}\:
and
Akihiro Ishibashi$^{2}$\thanks{\tt akihiro.ishibashi@kek.jp}\:
\\ \\
{\it ${}^{1}$School of Mathematics, Cardiff University} \\
{\it Cardiff, United Kingdom} \medskip \\
{\it ${}^{2}$KEK Theory Center,
Institute of Particle and Nuclear Studies
} \\
{\it High Energy Accelerator Research Organization (KEK)
} \\
{\it Tsukuba, Japan} \\
}

\date{02 February 2010}

\maketitle

\begin{abstract}
We place further restriction on the possible topology of stationary
asymptotically flat vacuum black holes in 5 spacetime dimensions.
We prove that the horizon manifold can be either a connected sum of Lens
spaces and ``handles''  $S^1 \times S^2$, or the quotient of $S^3$ by certain
finite groups of isometries (with no ``handles''). The resulting horizon
topologies include Prism manifolds and quotients of the Poincare homology
sphere. We also show that the topology of the domain of outer communication
is a cartesian product of the time direction with a finite connected sum
of $\mr^4,S^2 \times S^2$'s and $CP^2$'s, minus the black hole itself.
We do not assume the existence of any Killing vector beside
the asymptotically timelike one required by definition for stationarity.
\end{abstract}



\section{Introduction}

In this paper, we derive further restrictions on the possible topologies of 5-dimensional,
asymptotically flat, analytic, non-extremal, vacuum black hole spacetimes $(M,g)$
with compact horizon.
The known solutions in this class to date are the Myers-Perry black holes~\cite{mp}
(with horizon topology $S^3$), and the Emparan-Reall black rings and their
generalizations~\cite{er,senkov} (with horizon topology $S^2 \times S^1$). Whether these are all possible solutions
is unclear at present, but it has been conjectured by Reall that there could be solutions with only precisely one extra $U(1)$-Killing field, $\psi$. This conjecture has recently received some support by the investigations of~\cite{blackfold}.
Our aim is to place some limits on the possible topologies of such, as yet
conjectural, solutions, both in as far as the horizon is concerned, but also in as far as that of the entire domain of
outer communication (exterior of the black hole) in $M$ is concerned.

Several general theorems are already known in this direction, and our analysis is to a large extent
a combination of these: Firstly, it has been shown by~\cite{gs} (see also~\cite{gal,ra}) that the horizon cross section, $H$, can carry a metric of
strictly positive scalar curvature. This result, which holds in $D$ dimensions under the
assumption that matter satisfies the dominant energy condition, was proved originally in 4 dimensions
by Hawking \cite{Hawking}, where it implies that the horizon has topology $S^2$. It
implies strong restrictions on the horizon topology also in higher, especially 5-, dimensions.
Secondly, the topological censorship theorem~\cite{fsw,gc,wol} states that
any curve in the domain of outer communication with endpoints in the asymptotic region can
be deformed to a curve entirely within that region. Therefore, if the spacetime
is asymptotically flat in the standard sense, then the domain of outer communication is
simply connected. Thirdly, it is known that if the horizon is
rotating, i.e. if the original asymptotically timelike Killing field $t$ is not
tangent to the null generators of the horizon, then the rigidity theorem implies that there is at least one further
Killing field $\psi$ generating an action of $U(1)$ on spacetime which commutes with $t$ \cite{hiw,hiw1,im}.
If there is precisely one such $U(1)$ Killing field, then we have on the horizon,
\ben\label{kdef}
\xi = t + \Omega \psi \, ,
\een
where $\xi$ is tangent to the null generators of the horizon, and where the constant $\Omega$ is
the angular velocity of the horizon. In the known exact black hole solutions,
we have in fact two further extra Killing fields generating an action of $U(1) \times U(1)$ instead of
just $U(1)$. In that case, a complete classification of the possible solutions is available~\cite{hy,hy1}.
In particular, the possible topologies of $H$ are $L(p,q), S^2 \times S^1, S^3$. Furthermore, the
domain of outer communication then has topology $\mr \times \Sigma$, where $\Sigma$
can be shown using results of~\cite{orlik1} to be a direct sum of $\mr^4$, and copies of $S^2 \times S^2$, $\pm CP^2$'s,
minus the black hole $B$ itself.
Fourthly, if the horizon is non-rotating, then the solution is isometric
to the Schwarzschild spacetime~\cite{sud,gibbons,rogatko,ruback}, with horizon topology $S^3$, and
$\Sigma \cong \mr^4 \setminus B$. Finally, some papers have also appeared concerning the nature of
the past endpoint set of an event horizon in $D$-dimensions when the spacetime is dynamically evolving, see e.g.~\cite{ida}. However, these do not appear to give further constraints on the final
topology of the black hole beyond the ones that we have already mentioned.

\section{Results}

In this paper, we will consider the {\em generic} case of an analytic, stationary, rotating, vacuum black hole
with compact, connected\footnote{Our results can easily be generalized to multiple horizons.} 
horizon cross sections, which, as we have explained, might only have isometry group
$\mr \times U(1)$. As we will see, the statement about the topology of $\Sigma$ remains true in that case, but the possibilities for the horizon topology that we derive are more 
than just $L(p,q), S^2 \times S^1, S^3$. More precisely, we will prove the following two results:

\begin{res}
The topology of $H$ can be one of the following:

\begin{enumerate}

\item If $\psi$ has a zero on $H$, then the topology of $H$ must be\footnote{Here we allow that
the Lens space be $L(0,1):=S^3$.}
\ben\label{hdecomp1}
H \cong \# \, l \cdot (S^2 \times S^1) \, \# \, L(p_1, q_1) \, \# \cdots \# \, L(p_k, q_k) \, .
\een
Here, $k$ is the number of exceptional orbits of the action of $U(1)$ on $H$ that is generated by
$\psi$, and $l$ is the number of connected components of the zero set of $\psi$.

\item
If $\psi$ does not have a zero on $H$, then $H \cong S^3/\Gamma$, where $\Gamma$ can be
certain finite
subgroups of $SO(4)$, or $H \cong S^2 \times S^1$. This class of manifolds includes again the Lens-spaces,
but also Prism manifolds, the Poincare homology sphere, and various other
quotients. All manifolds in this class are certain Seifert fibred spaces over
$S^2$. The precise classification of the possibilities is given below
in table~\ref{tab1}.
\end{enumerate}
\end{res}

Thus, in summary, our first result is that $H$ is either a connected sum of handles and
Lens-spaces, or a certain kind of other spherical manifold with no handles. Our result is
somewhat stronger than what is implied by merely knowing that $H$ carries a
metric of positive scalar curvature, because it rules out the possibility that 
$H$ could be a connected sum of spherical manifolds. Our result is definitely much stronger
than what can be concluded from merely knowing that it carries a $U(1)$-action.

If the topology of $M$ is such that it allows for an
action of $U(1) \times U(1)$--as would be the case if the black hole solution could be connected continuously
to a solution with two commuting axial Killing fields rather than just one--then the possibilities for the topology of $H$ are cut down to
$L(p,q),S^2 \times S^1,S^3$, in particular, the other spherical manifolds,
such as Prism manifolds, cannot appear.

Our second result concerns the topology of the domain of outer communication.

\begin{res}
There is a compact manifold $B$ with boundary 
$\partial B = H$ (the ``black hole'') such that 
the domain of outer communication has topology
$M \cong \Sigma \times \mr$,
where
\ben\label{decomp1}
\Sigma \cong \Bigg( \mr^4 \, \# \, n \cdot (S^2 \times S^2) \, \# \, n' \cdot(\pm CP^2) \Bigg) \setminus B \, .
\een
for some $n,n'\in\mn$.
\end{res}
Thus, from the topological viewpoint, the condition that there be an isometry group
$\mr \times U(1) \times U(1)$ gives essentially the same restrictions on the topology of $\Sigma$
as just assuming stationarity. For the known black hole solutions we have $n=0=n'$, and
this may well be in fact the only possibility, although we cannot prove this. If we assume,
as is very reasonable, that $M$ carries a spin structure, then $n'=0$.

\begin{figure}[h]\begin{center}
\includegraphics[width=11cm]{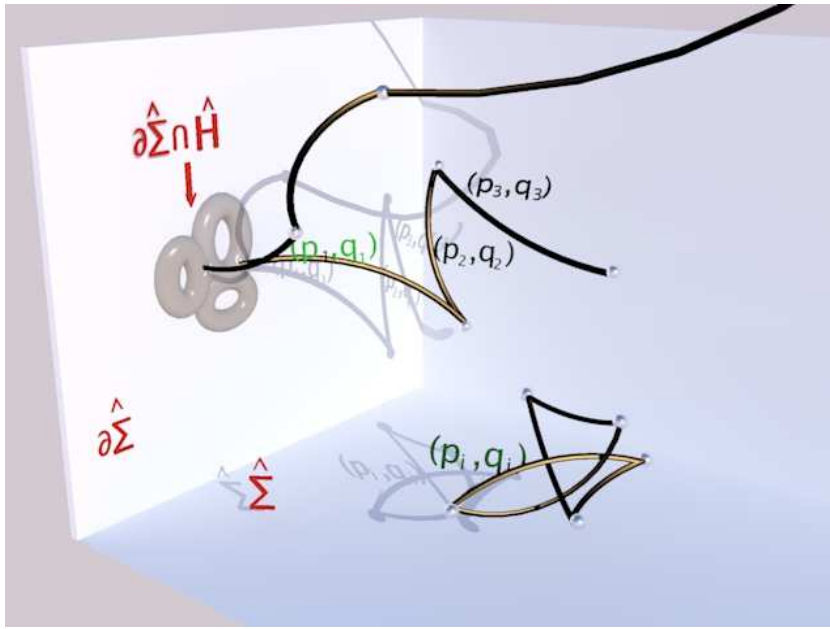} \hspace{1cm}
\caption{\label{fig5}
\small{The figure shows how the ``decorated orbit space'' $\hat \Sigma = \Sigma/U(1)$ can
look like. The weighted polyhedral arcs correspond to points in $\Sigma$ having a non trivial isotropy group $\mz_{p_{i}}\subset U(1)$. The boundary components correspond to points where the isotropy group is $U(1)$ (i.e. where $\psi=0$), or the horizon.
}}
\end{center}\end{figure}

Because $\Sigma$ is a manifold {\em together} with an action of $U(1)$, we can associate
further invariants with the solution that specify the
action. These invariants (considered first in~\cite{fin, fin2})
constitute the ``decorated orbit space'' $\hat \Sigma =
\Sigma/U(1)$, which consists of a manifold $\hat \Sigma$ with boundary together with
a collection of polyhedral arcs, the components of which are decorated by
certain integers, $(p_i,q_i)$, and Euler numbers, $e_i$, as in eq.~(\ref{Xdecor}). These data specify both the topology of $\Sigma$, as
well as the precise nature of the $U(1)$-action, see figure~\ref{fig5}.
(Note that the same spaces can carry different
inequivalent $U(1)$ actions.) Below in sec.~\ref{result2}, we also give conditions on the
decorated orbit space $\hat \Sigma$, under which the action of $U(1)$ on $M$ can be extended,
topologically, to an action of $U(1) \times U(1)$.
Of course, this is only a topological statement, the second factor of
$U(1)$ need not act by isometries. The point of this statement is that, if a solution
is continuously connected to a solution with the higher symmetry $U(1) \times U(1)$, i.e.
a finite perturbation, then one obtains some additional information about
the invariants in this way.

\medskip

\medskip

\subsection{Result 1}

\begin{figure}[h]
  \begin{center}
  \includegraphics[width=15cm]{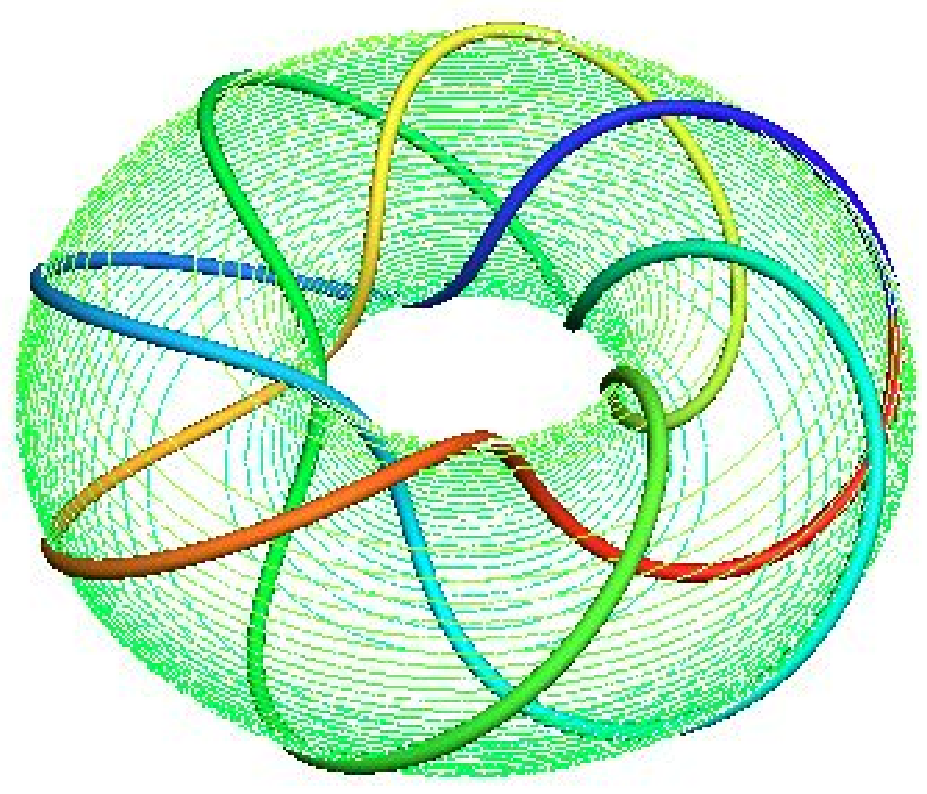} \hspace{1cm}
  \caption{\label{fig1}
   \small{A fibre with $p=3$, $n=7$} }
  \end{center}
 \end{figure}

\begin{figure}[h]
  \begin{center}
  \includegraphics[width=10cm]{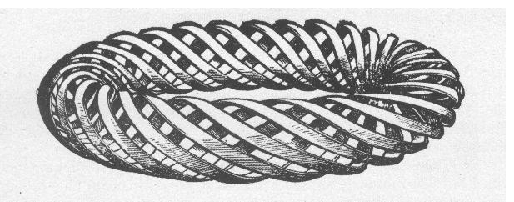} \hspace{1cm}
  \caption{\label{figTorus}
   \small{A more artistic version of a Seifert torus, from ``La Pratica della Perspectiva'' (1569),
   by D. Barbaro. }
  }
  \end{center}
\end{figure}

As we have explained, the rigidity theorem implies the existence of a second Killing field
$\psi$ commuting with $t$. The Killing field $\psi$ generates an action of $U(1)$ on the spacetime. 
The proof of this theorem~\cite{hiw,hiw1,im} implies that we can choose a horizon cross section $H$ 
such that $\psi$ is tangent to $H$, and such that $\psi$ is a Killing vector of the induced metric 
$\gamma$ on $H$. Thus, $(H,\gamma)$ is a compact Riemannian 3-manifold with an isometric action of $U(1)$. Let
\ben
\hat H = H/U(1)
\een
be the quotient space. As is well-known~\cite{orlik,raymond}, $\hat H$ is a 2-dimensional orbifold with boundary.
The boundary
\ben
\partial \hat H \cong \bigcup_{i=1}^l S^1_i
\een
consists of $l$ disjoint circles. Points in this boundary correspond to the fixed points of the $U(1)$-action, 
i.e. the places in $H$ where $\psi=0$.
Furthermore, each orbifold point $x_i \in \hat H$ is
labelled by a pair $(p_i, q_i)$, where $i=1,\dots,k$, of relatively prime integers satisfying
$0<q_i<p_i$. Near such a point $\hat H$ is
modelled upon the quotient $D^2_i/\mz_{p_i}$ of a disk by the cyclic group of order $p_i$
acting within the disk $D_i^2$ by phases $\exp(2\pi \sqrt{-1}/p_i)$. The points $x_i$ correspond
to the singular $U(1)$-fibres in $H$. Each such fibre is surrounded by a solid 3-dimensional torus $D^2_i \times
S^1$, and in such a solid torus, each $U(1)$-orbit winds around the disk $n_i$-times as it goes $p_i$-times
around the $S^1$-direction, see figures~\ref{fig1} and \ref{figTorus}
($q_i n_i \equiv 1 \, {\rm mod} \ p_i$). We are going to
distinguish the cases $\partial \hat H = \emptyset$ and $\partial \hat H \neq \emptyset$.

\medskip

\medskip



\subsubsection{Case (i): $\partial \hat H = \emptyset$}

This is the more interesting case and happens when $\psi \neq 0$ everywhere on $H$. Then, topologically,
since $\hat H$ inherits an orientation from the 3-form $\epsilon$ on $H$ given by $i_\psi \epsilon$,
both $\hat H$ and the fibres are oriented\footnote{In the terminology of Seifert manifolds, our
manifolds are of type ``$Oo$''.}. Thus, $H$ is a Seifert fibred space~\cite{orlik,seifert} characterized by the decoration data
\ben\label{symb}
H: \quad \{ g; b; (p_1, q_1), \dots, (p_k, q_k) \} \, ,
\een
where $g$ is the genus of the oriented compact Riemann surface $\hat H$ without boundary. $b$ is
an integer that characterizes the topology of the principal fibre bundle obtained after drilling out the
exceptional fibres and replacing them with regular ones.
The manifold $H$ is determined, as a manifold with $U(1)$-action, by the symbol~\eqref{symb}, but different symbols obtained from a given one by certain operations give rise to the same $U(1)$-manifold~\cite{orlik,raymond}. Symbols that may not be transformed into each other by such manipulations may still give rise to manifolds of the same topology of $H$, but the corresponding spaces will then
have inequivalent actions of $U(1)$.

Our aim is to show that
the orbifold Euler characteristic\footnote{This is an invariant of any 
Seifert manifold, i.e. it is unchanged under the manipulations mentioned above.} of $\hat H$ is positive, i.e. that
\ben
\chi_{\rm orbifold}(\hat H) := 2 - 2g - \sum_{i=1}^k \left( 1- \frac{1}{p_i} \right) > 0 \, .
\een
By standard results on Seifert 3-manifolds with positive orbifold
Euler characteristic [see e.g. table~4 of the review~\cite{lutz}; originial refs. include~\cite{seifert,scott}], this restricts the possible decoration data and topologies to the following ones [excluding $S^2 \times S^1$ which is included in case (ii)]:
In all cases we have $g=0,k\le 3$, and the possible
fibrations~\eqref{symb}, their corresponding 3-manifold $H$,
and fundamental groups are summarized in table~\ref{tab1},
\begin{table}[h]
{\small
\begin{tabular}{c|c|c}
Topology of $H$ & $U(1)$-Fibration & Fundamental Group\\
\hline\hline
$S^3$ & & 0\\
\hline
$L(b,1)$ & $\{b\}$ & $\mz_{|b|}$\\
\hline
$L(bp_1 + q_1, n)$ & $\{b; (p_1,q_1) \}$ & $\mz_{|bp_1+q_1|}$\\
& with $|bp_1+q_1|>1$ & \\
& and $n=p_1$ mod $bp_1+q_1$& \\
& with $0<n<bp_1+q_1$ & \\
\hline
$L(bp_1p_2+p_1q_2+q_2p_1,mp_2 - nq_2)$ & $\{b; (p_1,q_1), (p_2,q_2)\}$ & $\mz_{bp_1p_2+p_1q_2+q_2p_1}$\\
& with $mp_1 - n(bp_1+q_1)=1$& \\
\hline
$P(r)$ & $\{-1; (2,1), (2,1), (r,1)\}$ & $D_r^*$ \\
\hline
generalized Prism space& $\{b; (2,1), (2,1), (p_3,q_3)\}$ & $\mz_{|(b+1)p_3 + q_3|} \times D_{p_3}^*$ \\
& & if ${\rm g.c.d.}((b+1)p_3 + q_3, 2q_3)=1$\\
& & $\mz_{|m|} \times D'_{2^{k+2} p_3}$\\
& & for $(b+1)p_3 + q_3 = 2^k m$\\
\hline
$S^3/T^*$ & $\{-1;(2,1),(3,1),(3,1)\}$ & \\
\hline
generalized octahedral space& $\{b;(2,1),(3,q_2),(3,q_3)\}$ & $\mz_{|6b+3+2q_2+2q_3|} \times T^*$ \\
& & if ${\rm g.c.d.}(12,6b+3+2q_2+2q_3)=1$ \\
& & $\mz_{|m|} \times T'_{8 \cdot 3^{k+1}}$ \\
& & for $6b+3+2q_2+2q_3 = 3^k m$\\
\hline
$S^3/O^*$ & $\{-1;(2,1),(3,1),(4,1)\}$ & $O^*$\\
\hline
generalized cube space &$\{b;(2,1),(3,q_2),(4,q_3)\}$& $\mz_{|12b + 6 + 4q_2 + 3 q_3|} \times O^*$\\
\hline
$S^3/I^*$ & $\{-1;(2,1),(3,1),(5,1)\}$ & $I^*$ \\
\hline
generalized dodecahedral space & $\{b; (2,1), (3, q_2) (5, q_3)\}$ & $\mz_{|30b + 15 +10q_2 + 6q_3|} \times I^*$
\end{tabular}
\caption{The orientable Seifert manifolds with positive orbifold Euler characteristic}
\label{tab1}
}
\end{table}
where $D^*_n$ is the binary dihedral group of order $4n$, $T^*$ is the binary tetrahedral group
of order $24$, $O^*$ is the octahedral group of order $48$, $I^*$ the icosahedral group of order $120$
and $D'_{2^k(2n+1)}, T'_{8\cdot 3^k}$ are groups of the indicated order that are given e.g. in~\cite{lutz}.
In all cases, the fundamental group $\pi_1(H) = \Gamma$ is finite, and by
the Thurston elliptization
theorem for 3-manifolds (see e.g.~\cite{kleiner}),
it follows that $H = S^3/\Gamma$ in all cases.

Thus, what remains to be shown is that the orbifold Euler characteristic is
positive. As above, let $\gamma$ be the Riemannian metric on $H$ induced by the spacetime metric. Then, as shown by
\cite{gs}, there exists a positive function $\phi>0$ on $H$ such that conformally transformed metric
$\tilde \gamma = \phi^{-3/2} \gamma$ is a metric
with non-negative scalar curvature $\tilde S \ge 0$.
The metric $\gamma$ is invariant under $U(1)$,
so if we could show that also $\phi$ can be chosen $U(1)$ invariant, then also $\tilde \gamma$ is $U(1)$-invariant
and furthermore has positive scalar curvature. It is easily seen that the argument of~\cite{gs} can be
adapted in a straightforward way to prove this. For completeness, we indicate how this is done following~\cite{gs}.
Let us introduce Gaussian null-coordinates (see e.g.~\cite{im}) near the horizon as
\ben
g = 2 du( dv + r \, \beta_a dx^a + r \, \alpha du) + \gamma_{ab} dx^a dx^b \, ,
\een
where the indices $a,b, \dots$ indicate tensor components tangent to $H$. The Killing field $\xi$, see eq.~\eqref{kdef}, is given in these coordinates by $\xi = \partial/\partial u$, and the horizon is at $r=0$.
The function $\alpha$ is constant on
$H$ and given by the surface gravity of the black hole. By considering variations of
$H$ along an outward directed spatial normal vector field, it is demonstrated in~\cite{gs} that
there holds the inequality
\ben
\int_H \left( (\nabla^a f) \nabla_a f + \frac{1}{2} \{S - ({\mathscr L}_\xi \gamma)^{ab} ({\mathscr L}_\xi \gamma)_{ab}\} f^2 \right) \sqrt{\gamma} \, d^3 x \ge 0\, ,
\een
for any smooth function $f$ on $H$,
where indices are raised with $\gamma^{ab}$. Since $\xi$ is a Killing field, one can show that
the Lie-derivative in fact vanishes in our situation. In view of the inequality, one knows that the
spectrum $\{ \lambda_1, \lambda_2, \dots \}$
of the differential operator $-\nabla^a \nabla_a + \frac{1}{2} S$
is non-negative. It is then
possible, by standard results~\cite{kazdhan}, to  choose a strictly positive eigenfunction, $\phi>0$, for the first
eigenvalue $\lambda_1\ge 0$. The only additional
new observation necessary for us is that, since the differential operator commutes with the flow of $\psi$, we
may choose $\phi$ to be invariant as well. If it is not initially, we simply make it $U(1)$ invariant
by taking instead the average
\ben
\phi(x) \to \frac{1}{2\pi} \int_0^{2\pi} \phi \circ \theta_\tau(x) \, d\tau
\een
along the flow $\theta_\tau$ of $\psi$, which is again strictly positive everywhere on $H$ and
an eigenfunction of $-\nabla^a \nabla_a + \frac{1}{2} S$. The metric $\tilde \gamma = \phi^{-3/2} \gamma$
has non-negative scalar curvature $\tilde S$, because
\ben
\tilde S =  \phi^{-1}\left( 2\lambda_1 + \frac{3}{2\phi^2} (\nabla_a \phi)\nabla^a \phi \right) \ge 0 \, ,
\een
by the standard conformal transformation formula for the scalar curvature. This still leaves the possibility
that $\tilde S = 0$ everywhere on $H$. To rule out this case, one can argue as follows. Let $\tilde S_{ab}$ be the
Ricci tensor of $\tilde \gamma_{ab}$. Then, following Bourguignon (see~\cite{kazdhan}),
by considering deformations of $\tilde \gamma_{ab}$ in the direction of $\tilde S_{ab}$,
one could find a metric on $H$ which is Ricci flat, and since $H$ is a 3-manifold,
flat. The only possibility is then $H \cong T^3$, but this case has been ruled out by~\cite{gal}.

Thus, we can assume that $\tilde \gamma$ is $U(1)$-invariant and has non-negative scalar curvature $\tilde S \ge 0$
which is non-zero somewhere on $H$.
To continue, we recall that $H$ is a fibred space over $\hat H$, with fibres $S^1$, but it is not a principal
fibre bundle in the open neighborhoods of the exceptional fibres. We drill
out a neighborhood (solid 3-torus)
of the form $D_i^2 \times S^1$ around each exceptional fibre in $H$, where $D_i^2$ is a disk of
radius $r$ in Riemannian normal coordinates centered on the fibre. The resulting compact manifold with
boundary is denoted by $H_r$; its orbit space has the form
\ben
\hat H_r = \hat H \setminus \bigcup_{i=1}^k D^2_i \, ,
\een
i.e. it is a closed 2-manifold of genus $g$ with $k$ disks cut out, and hence has a boundary given by a union
of $k$ circles $S_i^1, i=1, \dots, k$. The 3-manifold $H_r$ now has only regular fibres, so it has the structure of
a principal fibre bundle over the 2-manifold $\hat H_r$ with boundary. We can then perform a ``Kaluza-Klein'' reduction of the
metric $\tilde \gamma$ in the usual way, i.e., we can write
\ben\label{decompgamma}
\tilde \gamma = \e^{\nu}(d\varphi + \omega_i dx^i)^2 + \e^{-\nu} h_{ij} dx^i dx^j \, ,
\een
where $\varphi$ is a $2\pi$-periodic coordinate on the fibres and $x^i$ are local coordinates of $\hat H_r$,
so that $\psi = \partial/\partial \varphi$,
where $\nu$ is a scalar field, $\omega$ a $U(1)$-connection, and $h$ a metric on $\hat H_r$.
Furthermore,
as a standard calculation shows,
the scalar curvature $\tilde S$ of $\tilde\gamma$ can be decomposed as
\ben
\e^{-\nu} \tilde S = \R - \frac{1}{4} \e^{2\nu} {\mathcal F}_{ij} {\mathcal F}^{ij} - \frac{1}{2} (\partial_i \nu) \partial^i \nu \, ,
\een
where $\mathcal F$ is the curvature of $\omega$, $\R$ is the scalar curvature of $h$, and
all indices are raised with $h$. We multiply this equation with $\sqrt{h} \, d^2 x$, the invariant integration element on $\hat H_r$,
and integrate, taking $r>0$ so small that $\tilde S > 0$ somewhere on $\hat H_r$. Then we get:
\ben
0< \int_{\hat H_r} \left( \e^{-\nu} \tilde S + \frac{1}{4}\e^{2\nu} \mathcal{F}_{ij} \mathcal{F}^{ij} + \frac{1}{2}
(\partial_i \nu) \partial^i \nu \right) \sqrt{h} \, d^2 x = \int_{\hat H_r} \R \sqrt{h} \, d^2 x
\, .
\een
On the right side, we now apply the Gauss-Bonnet theorem for the manifold with boundary $\hat H_r$. Letting
$K$ be the extrinsic curvature of the boundary components $S_i^1, i=1, \dots, k$ oriented by
the outward pointing normal, and $ds$ be the corresponding invariant line element, we get
\ben
0 < 2-2g - \sum_{i=1}^k \left( 1-\frac{1}{2\pi}\int_{S_i^1} K ds \right) \, .
\een
The remaining task is to evaluate the boundary integrals in the limit as $r \to 0$. For small $r$ and within the $i$-th
removed disk $D_i^2$, the metric $h$ takes
the form $h \sim dr^2 + r^2 dy^2$ up to higher orders of $r$, where $y$ is a coordinate which is
$2\pi/p_i$-periodic. This immediately gives the desired result for the orbifold Euler characteristic,
since the boundary integrals then evaluate to $1/p_i$ in the limit as $r \to 0$.

\subsubsection{Case (ii): $\partial \hat H \neq \emptyset$}

In this case, the orbit space $\hat H$ is a 2-dimensional oriented, compact orbifold with $l$ boundaries $S_i^1, i= 1, \dots, l$, and
$k$ orbifold points labelled by $(p_i, q_i), i=1, \dots, k$. Our first aim is to prove that
topologically
\ben\label{hdecomp2}
\hat H = D_0^2 \setminus \bigcup_{i=1}^l D_i^2 \, ,
\een
i.e. $\hat H$ is a large disk $D_0^2$, with $l$ small disks removed,
see figure~\ref{fig2}.
On this large disk, there are
$k$ orbifold points. To see this, we make use of the topological censorship theorem already
mentioned above. The point is that $H$ is the boundary of a spatial slice $\Sigma$.
As shown in~\cite{chrcosta}, we may choose this slice in such a way that it is invariant under the $U(1)$-action
(i.e., $\psi$ is tangent to $\Sigma$), and of course we have $\partial \Sigma = -H \cup S^3_\infty$, where
we mean a 3-sphere at infinity.

Then, by standard arguments, the quotient $\hat H$ must lie on the boundary of the corresponding
quotient $\hat \Sigma = \Sigma/U(1)$. The quotient of the 4-manifold $\Sigma$ by $U(1)$ is discussed in
more detail in the next subsection. Here we only need to know that $\hat \Sigma$ is
a space which locally is a manifold with boundaries, up to certain singularities that are localized
along 1-dimensional curve segments. One of the boundaries (that reaching out to infinity) has
topology $\mr^2$ outside a compact set, and $\hat H$ is a subset of this,
see figure~\ref{fig3}.
Now let us assume that instead of a disk $D_0^2$
we would have a disk $D_0^2$ with $h$ additional handles attached.
Then it is quite obvious, see figure~\ref{fig4} for an example,
that we could then find in $\hat \Sigma$ a curve that slings through one of these handles and is hence not
contractible. However, by the topological censorship theorem, in the domain of outer communication, any
curve is contractible, and therefore by standard topological arguments, so is any curve in $\hat \Sigma$.
Hence, we have a contradiction unless $h=0$.

\begin{figure}[h]\begin{center}
\include{orbsp2} \hspace{1cm}
\caption{\label{fig2}
\small{ We claim that the orbit space $\hat H$ is topologically
a disk with some disks removed. The crosses represent the orbifold points.
}}
\end{center}\end{figure}

\begin{figure}[h]\begin{center}
\includegraphics[width=10cm]{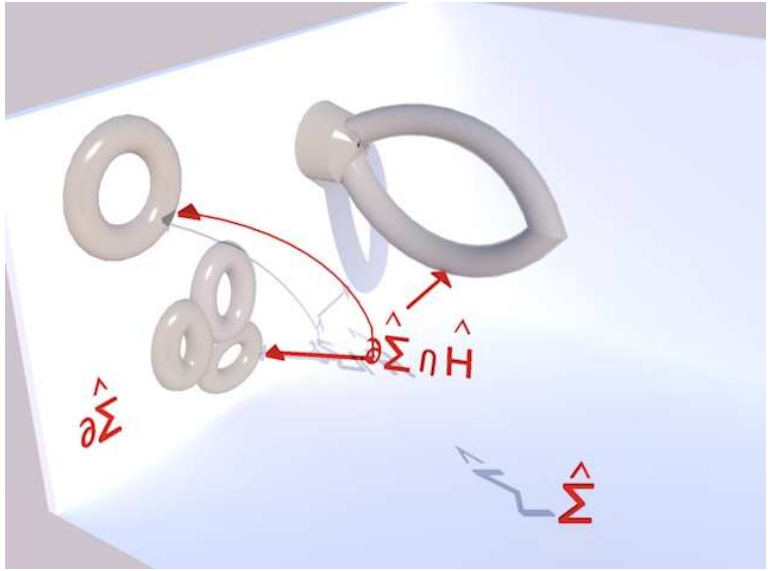} \hspace{1cm}
\caption{\label{fig3}
\small{This figure shows how the orbit space $\hat \Sigma$ looks like.
The orbit space $\hat H$ of the horizon forms part of the boundary of this space.
In fact, $\hat H$ should be connected, and we argue that there cannot be any handles
as suggested in this figure.
}}
\end{center}\end{figure}

\begin{figure}[h]\begin{center}
\includegraphics[width=10cm]{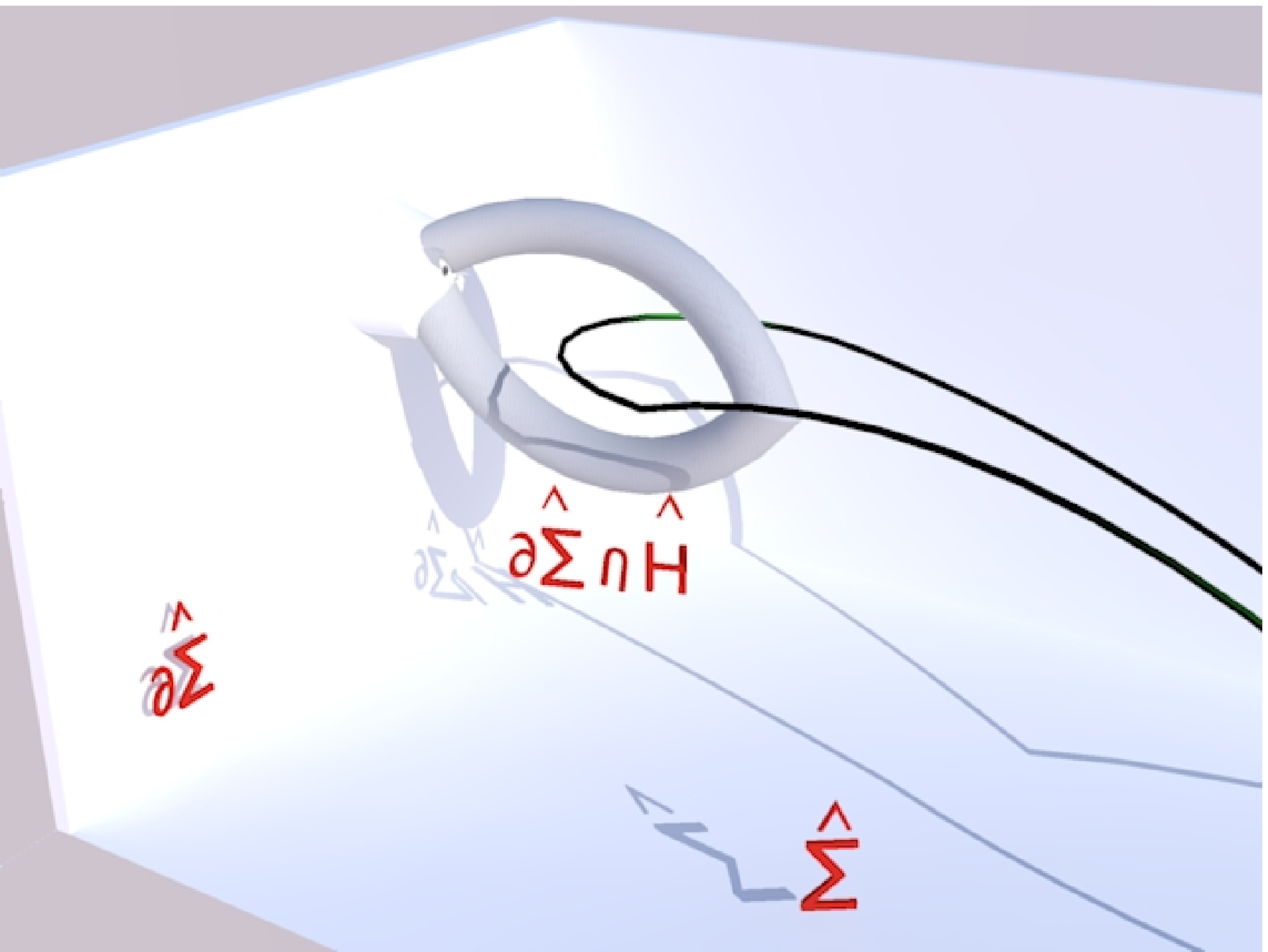} \hspace{1cm}
\caption{\label{fig4}
\small{If $\hat H$ contained a handle, then we could sling through it a curve as shown, and
this contradicts the topological censorship theorem.
}}
\end{center}\end{figure}

\begin{figure}[h]
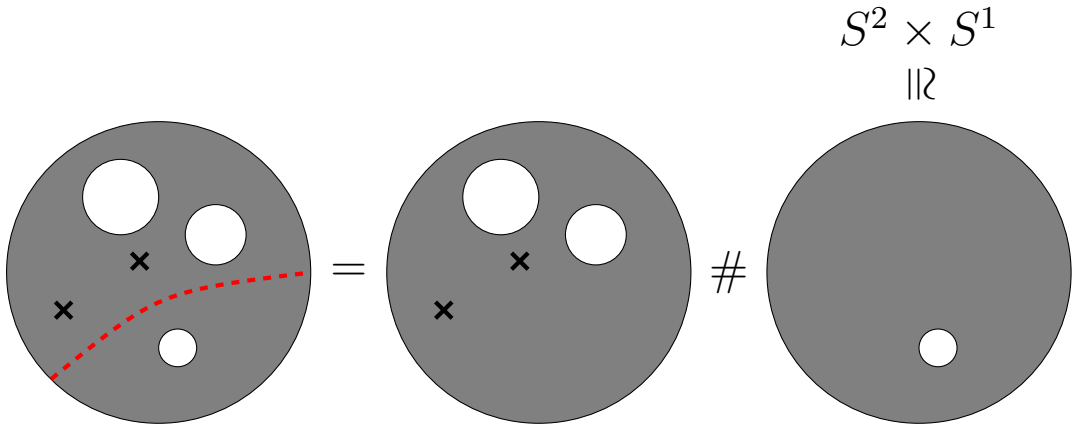
\begin{center}
\include{osd3}
\caption{\label{fig5a}
\small{ Removing a hole corresponds to removing a ``handle'' $S^2 \times S^1$.
}}
\end{center}\end{figure}

\begin{figure}[h]
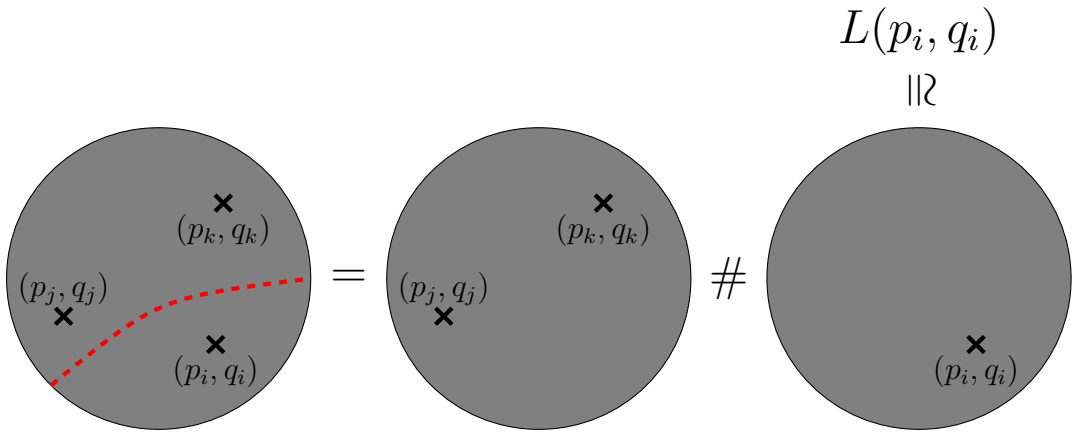
\begin{center}
\include{osd4}
\caption{\label{fig5b}
\small{Removing an orbifold point corresponds to removing a Lens space $L(p_i,q_i)$.
Here, $p_i=0$ is allowed (this corresponds to an $S^3$).
}}
\end{center}\end{figure}

The statement~\eqref{hdecomp2} now implies the desired decomposition eq.~\eqref{hdecomp1} by standard arguments of~\cite{orlik,raymond}.
For completeness, we briefly outline how these arguments are made. First, we cut out the removed
disks $D_i^2$, as illustrated in figure~\ref{fig5a}. Each of these operations corresponds, on the level of $H$, to
removing a handle $S^2 \times S^1$ and gluing back in a sphere. After removing $l$ such handles, we are left with a disk and $k$
orbifold points. These are now removed one by one, as illustrated in figure~\ref{fig5b}. Each of these operations corresponds, on the level of $H$,
to removing a Lens space $L(p_i,q_i)$ and gluing in a 3-sphere. Thus, we arrive at the desired
decomposition~\eqref{hdecomp1}. The condition that there exist a metric of positive scalar curvature
on $H$ does not give any further restrictions, since such decompositions are known to admit such metrics.

\medskip

\medskip

\medskip

\subsection{Result 2}
\label{result2}
We are going to prove result 2 by considering the orbit space $\hat \Sigma = \Sigma/U(1)$. For compact, simply connected
4-manifolds $X$ with a $U(1)$ action, the orbit space $\hat X=X/U(1)$ has been analyzed by~\cite{fin}.
He shows that the orbit space is a singular space which at generic points is a 3-manifold.
This 3-manifold has boundaries corresponding to certain fixed points of the action of $U(1)$,
together with certain piecewise smooth polygonal curves in $\hat X$, which
correspond to exceptional orbits where the isotropy group (i.e. the subgroup of $U(1)$ leaving a point invariant)
is discrete. More precisely, the nature of the orbit space is as follows: $\hat X = \hat L \cup \hat E \cup \hat F$,
where a hat always means the quotient by $U(1)$, and where $F$ is the space of fixed points in $X$ (where
the isotropy subgroup is $U(1)$),
$E$ is the space of exceptional orbits (where the isotropy group is $\mz_p \subset U(1)$ for some $p$), and
where $L$ is the set of regular orbits (where the isotropy subgroup is trivial).

\begin{enumerate}
\item\label{item1}
The set $\hat L$ is open in $\hat X$, and forms a smooth open manifold of dimension 3.

\item
The set $\hat F$ of fixed points is closed in $\hat X$. It consists of isolated points $x_i$, or
boundary components $\partial_i \hat X \cong S^2$. Near an isolated point, we can find a coordinate system
$(y_1, \dots, y_4)$ such that $x_i$ corresponds to the origin of the coordinate system, and such that
the action of an element $\e^{\sqrt{-1}t} \in U(1)$ is given by the matrix
\ben
\left(
\begin{matrix}
\cos p_i t & \sin p_i t & 0 & 0 \\
-\sin p_i t & \cos p_i t & 0 & 0\\
0 & 0 & \cos p_{i+1} t & \sin p_{i+1}t \\
0 & 0 & -\sin p_{i+1}t & \cos p_{i+1}t
\end{matrix}
\right) \, , \quad p_i, p_{i+1} \in \mz \, , \quad {\rm g.c.d.}(p_i, p_{i+1}) = 1 \, .
\een
Near each point of a boundary component, we can find a coordinate system
$(y_1, \dots, y_4)$ such that
the action of an element $\e^{\sqrt{-1}t} \in U(1)$ is given by the matrix
\ben
\left(
\begin{matrix}
1 & 0 & 0 & 0 \\
0 & 1 & 0 & 0\\
0 & 0 & \cos t & \sin t \\
0 & 0 & -\sin t & \cos t
\end{matrix}
\right) \, , 
\een
where $\{0 = r = \sqrt{y_1^2 + y_2^2} \}$ corresponds to the points in the boundary
component of $\hat X$, i.e. $(r, y_3, y_4), r>0$ provide coordinates for $\hat X$
near that boundary point.

\item \label{arcs}
The set $\hat E$ of exceptional orbits consists of smooth arcs $\gamma_i$ in $\hat X$. Each such arc is
labelled by a pair $(p_i, q_i)$ of integers. A point in $E$ has a neighborhood with coordinates
$(y_1, \dots y_4)$ such that the action of an element $\e^{\sqrt{-1} t} \in U(1)$ is given by the matrix
\ben
\left(
\begin{matrix}
1 & 0 & 0 & 0 \\
0 & 1 & 0 & 0\\
0 & 0 & \cos p_i t & \sin p_i t \\
0 & 0 & -\sin p_i t & \cos p_i t
\end{matrix}
\right) \, , 
\een
near such a point, the Killing field $\psi$ generating $U(1)$ is given locally
by $\partial/\partial y_1$, and the orbit space in the neighborhood of the arc
is parametrized by $y_2$, which runs along the arc, and $y_3,y_4$, running
transverse to the arc, located at $y_3=y_4=0$ locally.
Two arcs can intersect at an isolated fixed point (see previous item), and
the numbers $q_i \in \mz$ are then assigned in such a way that if the the adjacent arcs
carry $p_i$ and $p_{i+1}$, then we have
\ben\label{det}
\left|
\begin{matrix}
p_i & p_{i+1}\\
q_i & q_{i+1}
\end{matrix}
\right| = \pm 1 \, .
\een
\end{enumerate}
The final invariant associated with the $U(1)$ manifold $X$ comes from the Euler numbers of the
boundary components $\partial_i \hat X \cong S^2$ (if any). Let us shift the boundary
slightly inwards in $\hat X$. Then we obtain a surface denoted $S^2_i$ inside $\hat L$,
which is the base of a sub-$U(1)$ bundle in $X$ with base $S^2_i$ and fibres $U(1)$. We let
$e_i$ be the Euler ($=$ first Chern-) class of this bundle, i.e.
\ben\label{eulerclass}
e_i = \frac{1}{2\pi} \int_{S_i^2} {\mathcal F} \in \mz \, ,
\een
where $\mathcal F$ is the curvature of a connection in the $U(1)$-bundle over $S^2_i$ that can
be obtained by decomposing a metric on $\hat X$ similar to eq.~\eqref{decompgamma}.

It was shown in \cite{fin} that the above invariants
\ben\label{Xdecor}
X: \quad \{\hat X; e_1, \dots, e_b;  \gamma_1, (p_1, q_1), \dots, \gamma_k, (p_k, q_k) \}\, ,
\een
subject to the above constraint~(\ref{det}) are in one-to-one correspondence with the
compact, oriented, simply connected $U(1)$-manifolds $X$, i.e. for each set of invariants
there is precisely one such manifold, and vice-versa. Furthermore, it was shown in~\cite{fin, fin2} that the quadratic
form of $X$ (i.e., the pairing $Q_X:H_2(X) \times H_2(X) \to \mz$) is congruent over $\mz$ to the matrix

\ben
Q_X = \bigoplus m \cdot \left(
\begin{matrix}
0 & 1\\
1 & 0
\end{matrix} \right)  \oplus I_{m'} \oplus (-I_{m''})
\een
for some $m,m',m''\in \mn$. Such a quadratic form is obtained also for connected sums of copies of $\pm CP^2$ and copies of
$S^2 \times S^2$,
and therefore, since the topology of $X$ is uniquely determined by the
invariant $Q_X$ according to~\cite{freed,mil}, $X$ has to be topologically a connected sum of
copies of $S^2 \times S^2$, and $\pm CP^2$'s. The projective spaces
are forbidden if we assume, as appears to be reasonable from the physical viewpoint, that $X$ can carry a spin
structure.

In our case, we would like to take $X = \Sigma$, where $\Sigma$ is a spatial slice. We know that $\Sigma$ is simply connected by the
topological censorship theorem, but it is not compact. Its compactification is a manifold with boundary
$\partial \Sigma = -H \cup S^3_\infty$. Nevertheless, it is not difficult to generalize the
classification to this case. First, we glue a 4-dimensional ball $D^4_\infty$ into $\hat \Sigma$
along the boundary $\partial D^4_\infty = S^3_\infty$ at infinity, in such a way that the $U(1)$-actions
match up. We call the resulting manifold with boundary $\Sigma_{0}=\Sigma\cup D^{4}_{\infty}$. Then, if we take the quotient $\hat \Sigma_{0} = \Sigma_{0}/U(1)$, the quotient $\hat H$ will correspond to a (new) part of the boundary 
$\hat \Sigma_{0}$ that does not correspond to an axis as described in item \ref{item1}. above. 
The quotient of the horizon $H$ might additionally contain points in $\hat F$,
i.e. points corresponding to fixed points. Those correspond to
the following situation, see figure~\ref{fig5}.
\begin{enumerate}
\item
Let $x_i$ be a point in $\hat H \cap \hat F$ corresponding to an isolated fixed point. Then at the corresponding
points of $\Sigma_{0}$, we can choose coordinates $(y_1, y_2, y_3, y_4), y_1>0$ such that
the action of an element $\e^{\sqrt{-1}t} \in U(1)$ is given by the matrix
\ben
\left(
\begin{matrix}
1 & 0 & 0 & 0 \\
0 & 1 & 0 & 0\\
0 & 0 & \cos t & \sin t \\
0 & 0 & -\sin t & \cos t
\end{matrix}
\right) \, , 
\een
where $y_1 = 0$ locally corresponds to the boundary $H$. The quotient space $\hat \Sigma_{0}$
is locally parametrized by the coordinates $r = \sqrt{y_3^2 + y_4^2} >0, y_1>0$, and $y_2$, i.e.
it locally has the structure of a corner.
\item
Let $x_i$ be a point in $\hat H \cap \hat E$ corresponding to an exceptional orbit.
It has a neighborhood with coordinates
$(y_1, \dots y_4), y_1>0$ such that the action of an element $\e^{\sqrt{-1}t} \in U(1)$ is given by the matrix
\ben
\left(
\begin{matrix}
1 & 0 & 0 & 0 \\
0 & 1 & 0 & 0\\
0 & 0 & \cos p_i t & \sin p_i t \\
0 & 0 & -\sin p_i t & \cos p_i t
\end{matrix}
\right) \, . 
\een
Near such a point, the Killing field $\psi$ generating $U(1)$ is given locally
by $\partial/\partial y_2$, and the horizon is locally located at $y_1=0$.
The point $x_i$ corresponds to a singular fibre with labels $(p_i,q_i)$ in
$H$ (see above), where ${\rm g.c.d.}(p_i, q_i) = 1$. It is the intersection point of
$\hat H$ with one of the arcs in item \ref{arcs}. above.
\end{enumerate}
By a straightforward application of van Kampen's theorem, and using that
$\hat \Sigma_{0}$ is simply connected, the fundamental group of $\Sigma_{0}$ is then found to be
given by
\ben
\pi_1(\Sigma_{0}) = \mz_{p_1} \times \cdots \times \mz_{p_I}
\een
where $i=1,\dots,I$ runs through the set of isolated points on $\hat H$ which are connected to another such point
by an arc as in item \ref{arcs}. that is decorated by a pair $(p_i, q_i)$. Since we know that the fundamental group is in fact
trivial, and since by definition $p_i>0$, we can conclude that no such arc can exist in $\hat \Sigma_{0}$.

It is clear from this description that one can glue a decorated 3-manifold $\hat B$ with boundary
$\partial \hat B = \hat H$ into $\hat \Sigma_{0}$ so as to give a decorated 3-manifold $\hat X =\hat{\Sigma}_{0}\cup \hat{B}$ (see
eq.~\eqref{Xdecor}), and this will correspond, by the results of \cite{fin, fin2}, to a simply connected
four manifold $X$ with quadratic form $Q_X$ as above. Then the decomposition~\eqref{decomp1} follows, because for some $n,n'\in\mn$ \ben\Sigma=X\setminus (B\cup D^{4}_{\infty}) \cong \Bigg( \mr^4 \, \# \, n \cdot (S^2 \times S^2) \, \# \, n' \cdot(\pm CP^2) \Bigg) \setminus B \ . \een
An illustration of the weighted orbit space $\hat{\Sigma}$ is given above in figure \ref{fig5}. This space has as boundary components both points corresponding to the horizon ($\hat{H}$) as well as points corresponding to axes of the Killing field ($\partial \hat{\Sigma}\setminus \hat{H}$).

\begin{figure}[h]
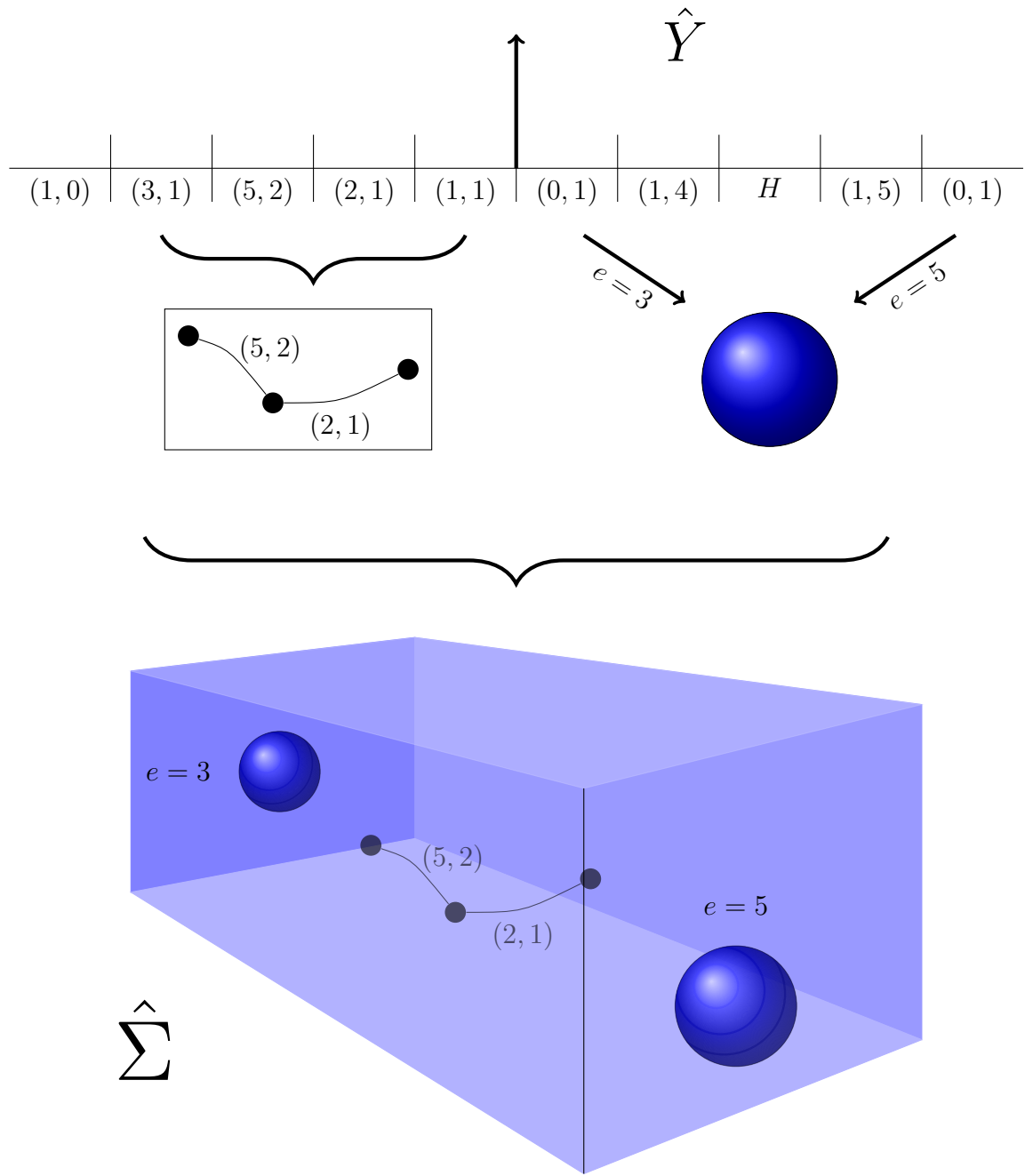
\begin{center}
\include{fig9}
\caption{\label{fig6}
\small{This figure shows the weighted 2-dimensional orbit space $\hat Y = \Sigma /[U(1) \times U(1)]$
and its relation to the corresponding 3-dimensional one $\hat \Sigma = \Sigma/[U(1) \times \{1\}]$.
Thus, if there are two $U(1)$-factors, the corresponding 3-dimensional weighted orbit space is
more restricted. In particular, in that case, there cannot be any knotted polyhedral arcs as suggested in figure \ref{fig5}
which represents the general situation.
}}
\end{center}\end{figure}

Finally, let us suppose that $(M,g)$ in fact has the isometry group $U(1) \times U(1)$ acting on the
spatial slice $\Sigma$. Then the decorated orbit space is in fact more restricted, as we shall now explain.
Let $\hat Y = \Sigma/[U(1) \times U(1)]$. Then, as shown in \cite{orlik1,hy}, the space $\hat Y$ is homeomorphic
to an upper half-plane, whose boundary is divided into several intervals $I_i, i=1, \dots , r$, labeled
by relatively prime integers $(p_i, q_i)$, except for a single special interval $I_h \cong H/[U(1) \times U(1)]$, see figure \ref{fig6}.
These integers specify which linear combination $p_i \psi_1+ q_i \psi_2=0$ of the two commuting
$U(1)$-Killing fields vanishes at the
points in $\Sigma$ corresponding to those in $I_i$ under the quotient by $U(1) \times U(1)$.
The first and last semi-infinite intervals are labeled, respectively,
by $(0,1)$ and $(1,0)$. This corresponds to the fact that, in the asymptotic region, the group action is
equivalent to that on $\mr^4$. If $I_{h-1},I_{h+1}$ are the intervals adjacent to the horizon interval, then
it is possible to see that $H \cong L(r,s)$, with
\ben\label{det1}
\left|
\begin{matrix}
p_{h-1} & p_{h+1}\\
q_{h-1} & q_{h+1}
\end{matrix}
\right| = r \, \quad sq_{h-1} = q_{h+1} + nr \, , \quad np_{h-1} = 1 \,\, {\rm mod} \,\, q_{h-1} \, .
\een
Now let $\hat \Sigma = \Sigma/[U(1) \times \{1\}]$
be the quotient by one $U(1)$-factor only. This gives a weighted orbit
space as described above consisting of a 3-manifold with boundaries and weighted arcs, see figure \ref{fig5}. The precise correspondence of this to $\hat Y$ is as follows.
As a space, $\Sigma \cong \mr^+ \times \mr^2 \setminus \cup_i D^3_i$, where
the boundary $S^2_i$ of the $i$-th ball corresponds to an interval $I_i$ that is
labeled by $(1,0)$. If $(p_{i-1},q_{i-1})$ resp. $(p_{i+1},q_{i+1})$ are the labels of the preceding resp.
following intervals, then it is possible to see that the $i$-th Euler class $e_i$ associated with $S^2_i$
[see eq.\eqref{eulerclass}] is given by
\ben\label{eulerclass1}
e_i = \frac{1}{2\pi} \int_{S_i^2} {\mathcal F} = \left|
\begin{matrix}
p_{i-1} & p_{i+1}\\
q_{i-1} & q_{i+1}
\end{matrix}
\right| \, .
\een
The last interval $(1,0)$ corresponds to the boundary component $\{0\}\times\mr^{2}$ of $\hat{\Sigma}$. The corresponding Euler class is found setting $p_{i+1}=0, q_{i+1}=1$.
The horizon $H$ corresponds to either a separate boundary sphere $S_h^2$,
or a part of $\{ 0 \} \times \mr^2$. The polyhedral arcs are obtained as follows.
Take the boundary $\partial \hat Y$ (a line), and delete any interval that is labeled by $(1,0),(0,1)$
together with its nearest neighbors,
and take away the interval $I_h$ corresponding to $H$.
This cuts the line into several connected pieces which are labeled each by a pair of
relatively prime integers. These connected pieces correspond to the polyhedral arcs.
The procedure is explained in figure~\ref{fig6}.

\medskip

\medskip

In summary, if the black hole has the isometry group $\mr \times U(1) \times U(1)$,
then the weighted orbit space of a slice is given by the simpler symbol
\ben\label{Xdecor1}
\Sigma: \quad \{\mr^+ \times \mr^2 \setminus \cup_i D^3_i; e_1, \dots, e_b;  \gamma_1, (p_1, q_1), \dots, \gamma_k, (p_k, q_k) \}\, ,
\een
where none of the polyhedral arcs close, and where the Euler classes are as above in eq.~\eqref{eulerclass1}.

\section{Outlook}

Let us finally discuss generalizations of our results. The most obvious question is whether one can obtain not only a classification of the topology,
but in fact of the metrics of stationary black holes. In the case
of symmetry group $ \mr \times U(1) \times U(1)$, this was achieved in \cite{hy,hy1}. There it was found
that what characterizes the solution uniquely (if it exists) are its conserved charges
as well as the data of the weighted orbit space $\hat Y$ (see previous section and
figure \ref{fig6}).
The decoration data include the collection of winding numbers $(p_i,q_i)$,
as well as the lengths $l_i$ of the intervals. In the situation considered
in this paper, one can only assume the existence of one $U(1)$, and the decorated orbit space
is now 3-dimensional (see eq. \eqref{Xdecor} and figure \ref{fig5}), and
has labelled polyhedral curves and certain areas on the boundary that correspond to the horizon. In this case, one would expect that one needs
further data to uniquely specify the spacetime metric, such as the length and angles between the arc segments, and/or e.g.
the areas of the horizon domain (the shaded area in figure \ref{fig5}). The latter has been suggested by~\cite{harmark}, but he seems to ignore the polyhedral arcs.

The second, easier,
question is what happens when matter fields are included. Our results rely on the rigidity theorem \cite{hiw,hiw1,im}
the topology theorem~\cite{gs}, and the topological censorship theorem~\cite{gc,wol}. For present proof of the rigidity theorem, the essential requirements are that the
null-energy condition holds, that the theory possess a well-posed (characteristic) initial value formulation, and that the domain of
outer communication be simply connected and analytic. These requirements hold
e.g. for a cosmological constant of either sign,
Maxwell fields, or the (bosonic sector of) minimal supergravity in 5 dimensions. The requirements of analyticity, initial value formulation, and simply connectedness (which
would fail e.g. for asymptotically Kaluza-Klein theories)
are actually only needed in order to define the extra $U(1)$-symmetry globally, but they are not required in order to merely conclude
that the horizon metric is $U(1)$-invariant. Similarly, the proof of~\cite{gs} only requires the dominant energy condition.
As a consequence, our result~1 will continue to hold for any Einstein-matter theory satisfying the dominant energy condition.
Result~2 on the other hand relies in an essential way on the fact that the domain of
outer communication is simply connected, which in turn is a consequence of the topological censorship theorem.
This theorem requires the null energy condition and allows one to conclude that the simply connectedness properties
of the spacetime are essentially the same as those of the asymptotic region. Therefore, this result will not generalize in the present form if e.g. the spacetime is asymptotically Kaluza-Klein, hence not simply connected. Also, result~2 relies in an essential way on the
global existence of a further $U(1)$ symmetry as guaranteed by the rigidity theorem. If the spacetime is not real analytic,
the present proofs do not work, and result~2 again does not seem to follow.

\medskip
{\bf Acknowledgements:} S.H. would like to thank the Institute of Particle and Nuclear
Studies, KEK, Japan, for its hospitality and financial support
in January 2010, during which time this work was carried out.

\bibliographystyle{utphys}
\bibliography{bhtop}

\end{document}

%% file: orbsp2.tex
\begin{tikzpicture}[scale=.75]
\filldraw[fill=gray] (0,0) circle (4cm);
\filldraw[fill=white] (1.5,1) circle (.8cm);
\filldraw[fill=white] (.5,-2) circle (.5cm);
\filldraw[fill=white] (-1,2) circle (1cm);
\node[scale=1.5] at (1.5,1) {$D_{i}^{2}$};
\path (-2.5,-1) node[scale=1.5] (a) {};
\path (2.4,-1.3) node[scale=1.5] (b){};
\node[scale=1.5] (c) at (-0.5,0.3) {};
\draw[ultra thick] (a.south west) -- (a.north east);
\draw[ultra thick] (a.north west) -- (a.south east);
\draw[ultra thick] (b.south west) -- (b.north east);
\draw[ultra thick] (b.north west) -- (b.south east);
\draw[ultra thick] (c.south west) -- (c.north east);
\draw[ultra thick] (c.north west) -- (c.south east);
\node[below, scale=1.5] at (a) {$(p_{i},q_{i})$};
\node[scale=2] at (-4.5,4.5) {$\hat{H}=H/U(1)$};
\node[scale=1.5] at (1.5,-4.5) {$D^{2}_{0}$};
\end{tikzpicture}

%% file: osd3.tex
\begin{tikzpicture}[scale=.5]
\filldraw[fill=gray] (0,0) circle (4cm);
\filldraw[fill=white] (1.5,1) circle (.8cm);
\filldraw[fill=white] (.5,-2) circle (.5cm);
\filldraw[fill=white] (-1,2) circle (1cm);
\path (-2.5,-1) node[scale=.75] (a) {};
\node[scale=.75] (b) at (-0.5,0.3) {};
\draw[ultra thick, red, dashed] (-2.828427,-2.828427) .. controls (0,-.5) .. (4,0) ;
\path (5,0) node[scale=1.5] {$=$};

\filldraw[fill=gray] (10,0) circle (4cm);
\filldraw[fill=white] (11.5,1) circle (.8cm);
\filldraw[fill=white] (9,2) circle (1cm);
\path (7.5,-1) node[scale=.75] (c) {};
\node[scale=.75] (d) at (9.5,0.3) {};
\path (15,0) node[scale=1.5] {$\#$};
\filldraw[fill=gray] (20,0) circle (4cm);
\filldraw[fill=white] (20.5,-2) circle (.5cm);
\node[rotate=270, scale=1.5] at (20, 5) {$\cong$};
\node[above, scale=1.5] at (20,5.5) {$S^{2}\times S^{1}$};

\draw[ultra thick] (a.south west) -- (a.north east);
\draw[ultra thick] (a.north west) -- (a.south east);
\draw[ultra thick] (b.south west) -- (b.north east);
\draw[ultra thick] (b.north west) -- (b.south east);
\draw[ultra thick] (c.south west) -- (c.north east);
\draw[ultra thick] (c.north west) -- (c.south east);
\draw[ultra thick] (d.south west) -- (d.north east);
\draw[ultra thick] (d.north west) -- (d.south east);

\end{tikzpicture}

%% file: osd4.tex
\begin{tikzpicture}[scale=.5]
\filldraw[fill=gray] (0,0) circle (4cm);
\path (-2.5,-1) node[scale=.75] (a) {};
\path (1.5,-1.75) node[scale=.75] (b) {};
\path (1.7,2) node[scale=.75] (c) {};
\node[above, scale=1] at (a) {$(p_{j},q_{j})$};
\node[below, scale=1] at (b) {$(p_{i},q_{i})$};
\node[below, scale=1] at (c) {$(p_{k},q_{k})$};
\draw[ultra thick, red, dashed] (-2.828427,-2.828427) .. controls (0,-.5) .. (4,0) ;
\path (5,0) node[scale=1.5] {$=$};

\filldraw[fill=gray] (10,0) circle (4cm);
\path (7.5,-1) node[scale=.75] (d) {};
\path (11.7,2) node[scale=.75] (f) {};
\node[above, scale=1] at (d) {$(p_{j},q_{j})$};
\node[below, scale=1] at (f) {$(p_{k},q_{k})$};
\path (15,0) node[scale=1.5] {$\#$};

\filldraw[fill=gray] (20,0) circle (4cm);
\path (21.5,-1.75) node[scale=.75] (e) {};
\node[below, scale=1] at (e) {$(p_{i},q_{i})$};
\node[rotate=270, scale=1.5] at (20, 5) {$\cong$};
\node[above, scale=1.5] at (20,5.5) {$L(p_{i},q_{i})$};

\draw[ultra thick] (a.south west) -- (a.north east);
\draw[ultra thick] (a.north west) -- (a.south east);
\draw[ultra thick] (b.south west) -- (b.north east);
\draw[ultra thick] (b.north west) -- (b.south east);
\draw[ultra thick] (c.south west) -- (c.north east);
\draw[ultra thick] (c.north west) -- (c.south east);
\draw[ultra thick] (d.south west) -- (d.north east);
\draw[ultra thick] (d.north west) -- (d.south east);
\draw[ultra thick] (e.south west) -- (e.north east);
\draw[ultra thick] (e.north west) -- (e.south east);
\draw[ultra thick] (f.south west) -- (f.north east);
\draw[ultra thick] (f.north west) -- (f.south east);

\end{tikzpicture}

%% file: fig9.tex
\begin{tikzpicture}[decoration=brace]
\draw (0,0) -- (15,0);
\foreach \x in {1.5,3,...,13.5}
\draw (\x,-.5) -- (\x,.5);
\node[below] at (.75,0) {$(1,0)$};
\node[below] at (2.25,0) {$(3,1)$};
\node[below] at (3.75,0) {$(5,2)$};
\node[below] at (5.25,0) {$(2,1)$};
\node[below] at (6.75,0) {$(1,1)$};
\node[below] at (8.25,0) {$(0,1)$};
\node[below] at (9.75,0) {$(1,4)$};
\node[below] at (11.25,0) {$H$};
\node[below] at (12.75,0) {$(1,5)$};
\node[below] at (14.25,0) {$(0,1)$};
\draw[ultra thick,->] (7.5,0) -- (7.5, 2);
\node[scale=2] at (10,2) {$\hat{Y}$};
\draw [ultra thick, decorate, decoration={brace, amplitude=.7cm}] (6.75,-1) -- (2.25,-1);
\node[scale=.8,circle, fill] (a) at (2.65, -2.5){};
\node[scale=.8,circle, fill] (b) at (3.9, -3.5){};
\node[scale=.8,circle, fill] (c) at (5.9, -3){};
\node[right] (d) at (3.255,-2.7) {$(5,2)$};
\node[below] (e) at (4.9, -3.5) {$(2,1)$};
\draw (a) .. controls (d.west) .. (b);
\draw (b) .. controls (e.north) ..(c);
\draw (2.3, -2.1) rectangle (6.25, -4.2);
\shadedraw [shading=ball] (11.25, -3.15) circle (1cm);
\draw[ultra thick,->] (8.5,-1) -- node[below, sloped] {$e=3$} (10, -2) ;
\draw[ultra thick,->] (14,-1) -- node[below, sloped] {$e=5$} (12.5, -2);

\filldraw[blue!50](1.8, -7.5) -- (6,-7) -- (6, -10) -- (1.8, -10.8) -- (1.8, -7.5);
\filldraw[blue!40](6.,-7) -- (13.5,-8) -- (13.5, -13) -- (6, -10) -- (6,-7);
\filldraw[blue!30](13.5,-13) -- (8.5,-15) -- (1.8, -10.8) -- (6, -10) -- (13.5,-13);
\filldraw[blue!20, opacity=0.4,](8.5,-9.25) -- (1.8,-7.5) -- (6, -7) -- (13.5, -8) -- (8.5,-9.25);
\draw (8.5,-15) -- (8.5, -9.25);

\shadedraw [shading=ball, opacity=.4] (4, -9) circle (.6cm);
\shadedraw [shading=ball, opacity=.4] (10.75, -12.5) circle (.9cm);
\node[scale=.8,circle, fill,opacity=.6] (a1) at (5.35, -10.1){};
\node[scale=.8,circle, fill,opacity=.6] (b1) at (6.6, -11.1){};
\node[scale=.8,circle, fill,opacity=.6] (c1) at (8.6, -10.6){};
\node[right,opacity=.6] (d1) at (5.955,-10.3) {$(5,2)$};
\node[below,opacity=.6] (e1) at (7.6, -11.1) {$(2,1)$};
\draw[opacity=.6] (a1) .. controls (d1.west) .. (b1);
\draw[opacity=.6] (b1) .. controls (e1.north) ..(c1);
\node[scale=3] at (2, -13) {$\hat{\Sigma}$};
\node at (10.75, -11){$e=5$};
\node at (2.5, -9){$e=3$};

\draw [ultra thick, decorate, decoration={brace, amplitude=.7cm}] (13,-5.5) -- (2,-5.5);

\end{tikzpicture}

%% file: hortop4.bbl
\providecommand{\href}[2]{#2}\begingroup\raggedright\begin{thebibliography}{10}

\bibitem{mp}
R.~C. Myers and M.~J. Perry, ``Black holes in higher dimensional space-times,''
\href{http://dx.doi.org/10.1016/0003-4916(86)90186-7}{{\em Ann. Phys.} {\bf
  172} (1986)  304}.

\bibitem{er}
R.~Emparan and H.~S. Reall, ``A rotating black ring in five dimensions,''
  \href{http://dx.doi.org/10.1103/PhysRevLett.88.101101}{{\em Phys. Rev. Lett.}
  {\bf 88} (2002)  101101},
\href{http://arxiv.org/abs/hep-th/0110260}{{\tt arXiv:hep-th/0110260}}.

\bibitem{senkov}
A.~A. Pomeransky and R.~A. Sen'kov, ``Black ring with two angular momenta,''
  \href{http://arxiv.org/abs/hep-th/0612005}{{\tt arXiv:hep-th/0612005}}.

\bibitem{blackfold}
R.~Emparan, T.~Harmark, V.~Niarchos, and N.~A. Obers, ``New horizons for black
  holes and branes,''
\href{http://arxiv.org/abs/0912.2352}{{\tt arXiv:0912.2352 [hep-th]}}.

\bibitem{gs}
G.~J. Galloway and R.~Schoen, ``A generalization of hawking's black hole
  topology theorem to higher dimensions,''
  \href{http://dx.doi.org/10.1007/s00220-006-0019-z}{{\em Commun. Math. Phys.}
  {\bf 266} (2006)  571--576},
\href{http://arxiv.org/abs/gr-qc/0509107}{{\tt arXiv:gr-qc/0509107}}.

\bibitem{gal}
G.~J. Galloway, ``Rigidity of outer horizons and the topology of black holes,''
\href{http://arxiv.org/abs/gr-qc/0608118}{{\tt arXiv:gr-qc/0608118}}.

\bibitem{ra}
I.~Racz, ``A simple proof of the recent generalisations of hawking's black hole
  topology theorem,'' \href{http://arxiv.org/abs/0806.4373}{{\tt
  arXiv:0806.4373}}.

\bibitem{Hawking}
S.~W. Hawking, ``Black holes in general relativity,''
\href{http://dx.doi.org/10.1007/BF01877517}{{\em Commun. Math. Phys.} {\bf 25}
  (1972)  152}.

\bibitem{fsw}
J.~L. Friedman, K.~Schleich, and D.~M. Witt, ``Topological censorship,''
  \href{http://dx.doi.org/10.1103/PhysRevLett.71.1486}{{\em Phys. Rev. Lett.}
  {\bf 71} (1993)  1486},
\href{http://arxiv.org/abs/gr-qc/9305017}{{\tt arXiv:gr-qc/9305017}}.

\bibitem{gc}
P.~T. Chrusciel, G.~J. Galloway, and D.~Solis, ``Topological censorship for
  kaluza-klein space-times,''
  \href{http://dx.doi.org/10.1007/s00023-009-0005-z}{{\em Annales Henri
  Poincare} {\bf 10} (2009)  893--912},
\href{http://arxiv.org/abs/0808.3233}{{\tt arXiv:0808.3233 [gr-qc]}}.

\bibitem{wol}
G.~J. Galloway, K.~Schleich, D.~Witt, and E.~Woolgar, ``The ads/cft
  correspondence conjecture and topological censorship,''
  \href{http://dx.doi.org/10.1016/S0370-2693(01)00335-5}{{\em Phys. Lett.} {\bf
  B505} (2001)  255--262},
\href{http://arxiv.org/abs/hep-th/9912119}{{\tt arXiv:hep-th/9912119}}.

\bibitem{hiw}
S.~Hollands, A.~Ishibashi, and R.~M. Wald, ``A higher dimensional stationary
  rotating black hole must be axisymmetric,''
  \href{http://dx.doi.org/10.1007/s00220-007-0216-4}{{\em Commun. Math. Phys.}
  {\bf 271} (2007)  699--722},
\href{http://arxiv.org/abs/gr-qc/0605106}{{\tt arXiv:gr-qc/0605106}}.

\bibitem{hiw1}
S.~Hollands and A.~Ishibashi, ``On the `stationary implies axisymmetric'
  theorem for extremal black holes in higher dimensions,''
  \href{http://dx.doi.org/10.1007/s00220-009-0841-1}{{\em Commun. Math. Phys.}
  {\bf 291} (2009)  403--441},
\href{http://arxiv.org/abs/0809.2659}{{\tt arXiv:0809.2659 [gr-qc]}}.

\bibitem{im}
V.~Moncrief and J.~Isenberg, ``Symmetries of higher dimensional black holes,''
  \href{http://dx.doi.org/10.1088/0264-9381/25/19/195015}{{\em Class. Quant.
  Grav.} {\bf 25} (2008)  195015},
\href{http://arxiv.org/abs/0805.1451}{{\tt arXiv:0805.1451 [gr-qc]}}.

\bibitem{hy}
S.~Hollands and S.~Yazadjiev, ``Uniqueness theorem for 5-dimensional black
  holes with two axial killing fields,''
  \href{http://dx.doi.org/10.1007/s00220-008-0516-3}{{\em Commun. Math. Phys.}
  {\bf 283} (2008)  749--768},
\href{http://arxiv.org/abs/0707.2775}{{\tt arXiv:0707.2775 [gr-qc]}}.

\bibitem{hy1}
S.~Hollands and S.~Yazadjiev, ``A uniqueness theorem for stationary
  kaluza-klein black holes,''
\href{http://arxiv.org/abs/0812.3036}{{\tt arXiv:0812.3036 [gr-qc]}}.

\bibitem{orlik1}
P.~Orlik and F.~Raymond, ``Actions of the torus on 4-manifolds. i,'' {\em
  Transactions of the American Mathematical Society} {\bf 152} (1970) no.~2,
  531. \url{http://www.jstor.org/stable/1995586}.

\bibitem{sud}
D.~Sudarsky and R.~M. Wald, ``Mass formulas for stationary einstein yang-mills
  black holes and a simple proof of two staticity theorems,''
  \href{http://dx.doi.org/10.1103/PhysRevD.47.R5209}{{\em Phys. Rev.} {\bf D47}
  (1993)  5209--5213},
\href{http://arxiv.org/abs/gr-qc/9305023}{{\tt arXiv:gr-qc/9305023}}.

\bibitem{gibbons}
G.~W. Gibbons, D.~Ida, and T.~Shiromizu, ``Uniqueness of (dilatonic) charged
  black holes and black p- branes in higher dimensions,''
  \href{http://dx.doi.org/10.1103/PhysRevD.66.044010}{{\em Phys. Rev.} {\bf
  D66} (2002)  044010},
\href{http://arxiv.org/abs/hep-th/0206136}{{\tt arXiv:hep-th/0206136}}.

\bibitem{rogatko}
M.~Rogatko, ``Uniqueness theorem of static degenerate and non-degenerate
  charged black holes in higher dimensions,''
  \href{http://dx.doi.org/10.1103/PhysRevD.67.084025}{{\em Phys. Rev.} {\bf
  D67} (2003)  084025},
\href{http://arxiv.org/abs/hep-th/0302091}{{\tt arXiv:hep-th/0302091}}.

\bibitem{ruback}
P.~Ruback, ``A new uniqueness theorem for charged black holes,'' {\em Classical
  and Quantum Gravity} {\bf 5} (1988)  L155.

\bibitem{ida}
D.~Ida and M.~Siino, ``Topology change of black holes,''
  \href{http://dx.doi.org/10.1143/PTP.118.715}{{\em Prog. Theor. Phys.} {\bf
  118} (2007)  715},
\href{http://arxiv.org/abs/0704.0100}{{\tt arXiv:0704.0100 [gr-qc]}}.

\bibitem{fin}
R.~Fintushel, ``Classification of circle actions on 4-manifolds,'' {\em
  Transactions of the American Mathematical Society} {\bf 242} (1978)  377.
  \url{http://www.jstor.org/stable/1997745}.

\bibitem{fin2}
R.~Fintushel, ``Circle actions on simply connected 4-manifolds,'' {\em
  Transactions of the American Mathematical Society} {\bf 230} (1977)  147.
  \url{http://www.jstor.org/stable/1997715}.

\bibitem{orlik}
P.~Orlik, {\em Seifert manifolds}.
\newblock Lecture Notes in Mathematics, Vol. 291. Springer-Verlag, Berlin,
  1972.

\bibitem{raymond}
F.~Raymond, ``Classification of the actions of the circle on 3-manifolds,''
  {\em Transactions of the American Mathematical Society} {\bf 131} (1968)
  no.~1, 51. \url{http://www.jstor.org/stable/1994680}.

\bibitem{seifert}
W.~Threlfall and H.~Seifert, ``Topologische untersuchung der
  diskontinuit{\"a}tsbereiche endlicher bewegungsgruppen des dreidimensionalen
  sph{\"a}rischen raumes (schlu{\ss}),'' {\em Mathematische Annalen} {\bf 107}
  (1933) no.~1, 543.

\bibitem{lutz}
F.~H. Lutz, ``Triangulated manifolds with few vertices: Geometric
  3-manifolds,'' \href{http://arxiv.org/abs/math/0311116}{{\tt
  arXiv:math/0311116}}.

\bibitem{scott}
P.~Scott, ``The geometries of 3-manifolds,'' {\em Bulletin of the London
  Mathematical Society} {\bf 15} (1983) no.~5, 401.

\bibitem{kleiner}
B.~Kleiner and J.~Lott, ``Notes on perelman's papers,''
  \href{http://arxiv.org/abs/math/0605667}{{\tt arXiv:math/0605667}}.

\bibitem{kazdhan}
J.~Kazdan and F.~Warner, ``Prescribing curvatures,'' {\em Proceedings of
  Symposia in Pure Mathematics} {\bf 27} (1975)  309.

\bibitem{chrcosta}
P.~T. Chrusciel and J.~Lopes~Costa, ``On uniqueness of stationary vacuum black
  holes,''
\href{http://arxiv.org/abs/0806.0016}{{\tt arXiv:0806.0016 [gr-qc]}}.

\bibitem{freed}
M.~Freedman, ``The topology of four-dimensional manifolds,'' {\em J.
  Differential Geom} {\bf 17} (1982) no.~3, 357.

\bibitem{mil}
J.~Milnor, {\em On simply connected 4-manifolds}.
\newblock 1958 Symposium internacional de topolog{\i}a algebraica.

\bibitem{harmark}
T.~Harmark, ``Domain structure of black hole space-times,''
  \href{http://dx.doi.org/10.1103/PhysRevD.80.024019}{{\em Phys. Rev.} {\bf
  D80} (2009)  024019},
\href{http://arxiv.org/abs/0904.4246}{{\tt arXiv:0904.4246 [hep-th]}}.

\end{thebibliography}\endgroup
